\documentclass[a4paper,fleqn,useAMS,usenatbib]{mnras}

\pdfoutput=1

\usepackage{mathptmx}

\usepackage[T1]{fontenc}
\usepackage{ae,aecompl}

\usepackage[dvips]{graphicx}
\usepackage{amsmath}
\usepackage{amsfonts}
\usepackage{amssymb}
\usepackage[dvipsnames]{xcolor}

\usepackage{hyperref}
\hypersetup{
	colorlinks=true,	
	urlcolor=MidnightBlue,		
    draft=false,
	hypertexnames=true,
	plainpages=false,
	naturalnames=true,
	pdftitle={The Weird Detector},    
	pdfauthor={Wheeler \& Kipping},     
	linkcolor=WildStrawberry,          
	citecolor=ForestGreen,        
}

\newcommand{\kepler}{{\it Kepler}}
\newcommand{\tess}{{TESS}}

\newcommand{\batman}{{\tt batman}}
\newcommand{\multi}{{\sc MultiNest}}
\newcommand{\sebastien}{{\tt weirddetector}}
\newcommand{\datalink}{\label{datalink}\href{https://github.com/CoolWorlds/WDdata}{this URL}\footnote{\url{https://github.com/CoolWorlds/WDdata}}}
\newcommand{\sebastienlink}{\href{https://github.com/CoolWorlds/weirddetector}{this URL}}

\newcommand{\revision}[1]{{#1}}
\newcommand{\lrevision}[1]{{#1}}

\title[The Weird Detector]{
The Weird Detector: Flagging periodic, coherent signals
of arbitrary shape in time series photometry
}
\author[Wheeler \& Kipping]{Adam Wheeler$^{1}$\thanks{E-mail:
\href{mailto:a.wheeler@columbia.edu}{a.wheeler@columbia.edu}} and David Kipping$^{1}$\\
$^{1}$Dept. of Astronomy, Columbia University, 550 W 120th Street, New York NY 10027}

\date{Accepted . Received ; in original form }

\pubyear{2018}

\begin{document}
\label{firstpage}
\pagerange{\pageref{firstpage}--\pageref{lastpage}}
\maketitle

\begin{abstract}
    By design, model-based approaches for flagging transiting exoplanets in light curves, such as boxed least squares, excel at detecting planets with low S/N at the expense of finding signals that are not well described by the assumed model, such as self-lensing binaries, disintegrating or evaporating planets, or planets with large rings.  So far, such signals have typically been found through visual searches by professional or citizen scientists, or by inspection of the photometric power-spectra. We present a nonparametric detection algorithm, for short duty-cycle periodic signals in photometric time series based on phase dispersion minimization. We apply our code to 161,786 \kepler\ sources and detect 18 new periodic signals consistent with heartbeat binaries/planets, 4 new singly-transiting systems, and 2 new doubly-transiting systems. We show that our code is able to recover the majority of known \kepler\ objects of interest (KOIs) to high confidence, as well as more unusual events such as Boyajian's star and a comet passing through the \kepler\ field. \revision{Nonparametric signal-flagging techniques, such as the one presented here,} will become increasingly valuable with the coming data from \tess\ and future transit surveys as the volume of data available to us exceeds that which can be feasibly examined manually.
\end{abstract}


\section{Introduction}
\label{sec:intro}

The photometric transit method has emerged as the most successful technique for
detecting new exoplanets. At the time of writing, the \textit{Kepler Mission}
alone has brought in a haul of over 4500 transiting planet candidates, of which
the majority have now been confirmed (see the NASA Exoplanet Archive;
\citealt{akeson:2013}). With \tess\ expected to detect some $10^4$ new examples
\citep{sullivan:2015,bouma:2017,barclay:2018,ballard:2018,huang:2018} and
future surveys looking set to pull in even larger yields (e.g. see
\citealt{rauer:2014,jacklin:2015,cortes:2018}), the transit method looks to
remain a critical tool to exoplanetary science for many years to come.

The algorithm used to detect new transiting planet candidates
varies between missions and teams. The most frequently cited
method is the Boxed Least Squares (BLS) algorithm \citep{kovacs:2002}, which
behaves as an optimal detector when the signal is i) strictly-periodic
ii) box-shaped iii) in the presence of Gaussian noise. Real
transits show limb darkening curvature \citep{knutson:2007} and
time-integration smoothing \citep{binning:2010}, but BLS remains nearly optimal
even when the light curve becomes somewhat U-shaped. Many alternative
algorithms have been proposed to BLS, which are usually designed to tackle
cases where either assumption i) or iii) are relaxed. As far as we can tell,
no transit detection algorithms have been explicitly designed with the
objective of relaxing assumption ii), however.

As an example, the Quasi-Periodic Automated Transit Search Algorithm (QATS)
assumes that transits are trapezoidal-like and in the presence of Gaussian
noise, but can be non-periodic \citep{carter:2013}. As another example,
time-correlated noise structure has been known to affect transit detection
since \citet{pont:2006}, and is most commonly tackled by pre-whitening and
detrending of the time series prior to searching (e.g. \citealt{tamuz:2005,
kovacs:2005,jenkins:2010,guterman:2015}), although recently deep learning
approaches have attacked the problem without this filtering step (e.g.
see \citealt{pearson:2018,shallue:2018,zucker:2018}). In all of the above,
an underlying assumption is that transits are expected to have a shape
consistent with a trapezoid.

Irregularly shaped transits, including asymmetric and multi-dip events, can
emerge in a variety of plausible astrophysical scenarios. Planetary rings
\citep{barnes:2004,zuluaga:2015} and moons \citep{luna:2011} are examples
we'd expect based on the Solar System planets, causing transits with an 
irregular morphology due to the overlap with another occulter. Disintegrating
planets \citep{rappaport:2012}, photoevaporation \citep{vidal:2003},
circumstellar material \citep{vanderburg15}, proto-satellite disks
\citep{mamajek:2012}, bow shocks \citep{llama:2013}, gravity darkening
\citep{barnes:2009}, planetary oblateness \citep{searger:2002}, atmospheric
refraction \citep{hui:2002}, and even extreme orbital eccentricities
\citep{kipping:2008} have all been argued to be other plausible astrophysical
effects which could distort the transit. Distorted transits have even been
argued to be a possible means of detecting advanced civilizations
\citep{arnold:2005,korpela:2015,cloaking:2016}. While small perturbations
to the transit should not greatly impact BLS's sensitivity, highly
irregular transits require another approach. 
Although there are presently only a
handful of known examples of these ``weird'' transits, with perhaps the most
dramatic being that of Boyajian's star \citep{boyajian:2016}, they represent some
of the most scientifically rich objects to date.

Arguably the most successful approach to date for searching for weird transits
has been through the citizen science Planet Hunters program
\citep{fischer:2012}. This program demonstrates that human beings certainly
have the ability to successfully identify unusual shaped signals, although any
approach using humans will face challenges with scalability and statistical
testing. An software-based solution could overcome such hurdles although to
date there has been little attention devoted to this issue. In this work,
we therefore aim to address this problem by presenting a so-called weird
detector for photometric time series.

\revision{
Our method is related to phase dispersion minimization (PDM), variants of which
have been employed for decades in other astronomical contexts.
To the best of out knowledge, a variant of PDM appeared first in 
\citet{whittiker:1926}, a general work on data analysis.
A similar idea was employed in \citet{lafler:1965}, which introduced the used of
``string-length'' methods, in their case to compute the periods of RR Lyrae 
stars.
Similar methods were developed again in \citet{jurkevich:1971}, 
\citet{warner:1972}, and \citet{stellingwerf:1978}, from which the term
``phase dispersion minimization'' arises.
Such methods used to supersede the discrete Fourier transform and 
the Lomb-Scargle periodogram \citep{lomb:1976, scargle:1982} for determining
the period of variable stars and stellar rotation because they do not require
the evaluation of trigonometric functions and were thus less expensive to run
given the limited computational resources of the time \citep{kovacs:2002}.
More recently, \cite{plavchan:2008} applied a similar algorithm to find both 
variable stars
and transiting exoplanets in data from the Two Micron All Sky Survey (2MASS)
\citep{skrutskie:2006}. \citet{parks:2014} also applies a ``binless'' PDM
variant to search for variable stars in 2MASS data. 
}

In Section~\ref{sec:methods}, we describe our algorithmic solution (dubbed
\sebastien; Julia implementation available at \sebastienlink) and the
details behind the code's operation and assumptions. In
Section~\ref{sec:results}, we explore the performance of our code in
application to the \kepler\ data, including numerous new detections. Finally,
we discuss the scalability, future improvements and applications of the weird
detector in Section~\ref{sec:discussion}.

\section{Methodology}
\label{sec:methods}

\subsection{Principles and Assumptions}

In designing an algorithm to detect irregular signals, one is faced with the
fundamental question of what one wishes to optimize for. In a least squares
regression problem, for example, the cost function is the
sum of the residuals (optionally weighted by their uncertainty) squared,
where the residuals are computed against some parametric model. In the case of
BLS, which is an example of a least squares problem, the model employed is a
periodic boxcar function defined by a period, phase, depth and duration
\citep{kovacs:2002}. \revision{For ordinary transits, this is a well-motivated 
simplification, but for signals with morphology deviating strongly from 
an inverted boxcar transit, the mismatch between model and data is problematic.}

One could imagine using a variant of the BLS code based on nested transits for
detecting rings and moons, or a version incorporating a skewed transit model
in order to cope with disintegrating planets. But each code would struggle with
other effects and both would be bespoke algorithms rather than generalized
weird detectors. We quickly decided that proposing any kind of parametric
function to describe the transit morphology would be limiting, in that our code
could then only detect signals which could be described by said function.
Ultimately, we wanted an algorithm which could handle the unexpected and even
have the ability to detect signals previously unimagined. At the same time,
the problem of detecting any and all effects optimally is both ill-defined
and formally intractable. Some underlying assumptions are always necessary.

In this work, our formal assumptions are that the weird signals we seek are
strictly periodic and in the presence of Gaussian noise - i.e. assumptions
i) and iii) used by BLS. We make only minimal assumptions about the signal's
shape. By virtue of the strict periodicity assumption, the signal is implicitly
not time varying and thus repeats each epoch. 
We highlight that ordinary planetary transits also satisfy this condition and
produce coherent signals in folded photometry, but more generally any repeating
signal will do so, be it a lensed black hole or an orbiting alien
megastructure. \revision{Boyajian's star satisfies these assumptions only
approximately, which means that we are limited to detecting its strictly
periodic analogs.}

\revision{We therefore loosely state that our objective function should identify signals
that are both coherent and have a transit-like morphology, in the sense of being
narrow, rather than extended across an entire phase. This is primarily to avoid
flagging brightness modulation stellar from rotation in the presence of starspots.}
Such signals should be associated
with high power, whereas pure Gaussian noise would produce low or ideally
negligible power. It should be noted that already at this stage, it can be
stated that \sebastien\  must have lower sensitivity to planetary transits than
BLS. This is because BLS includes additional information about the signal that
our algorithm does not - the shape of the expected events.  \revision{Because of
\sebastien's much weaker assumptions about signal shape, it has inferior
sensitivity for box-like signals. This is a necessary trade-off for any weird
detector though, imposed by the differing cost functions.}

\subsection{Quantifying Coherence and Morphology}

Having established that we wish to detect coherent signals, the next
step is to identify an approach for achieving this goal. Let us assume
that the number of epochs constituting a folded signal is large and
that any nuisance signals (e.g. stellar variability) have been removed.
We would expect that each epoch's phase curve is self-similar to the
others. In other words, the dispersion at any phase location is small.
We might imagine, then, that if we folded our time series on many trial
periods we could identify the correct period by seeing which one exhibits
the smallest phase dispersion.

\begin{figure}
    \centering
    \includegraphics[width=0.47\textwidth]{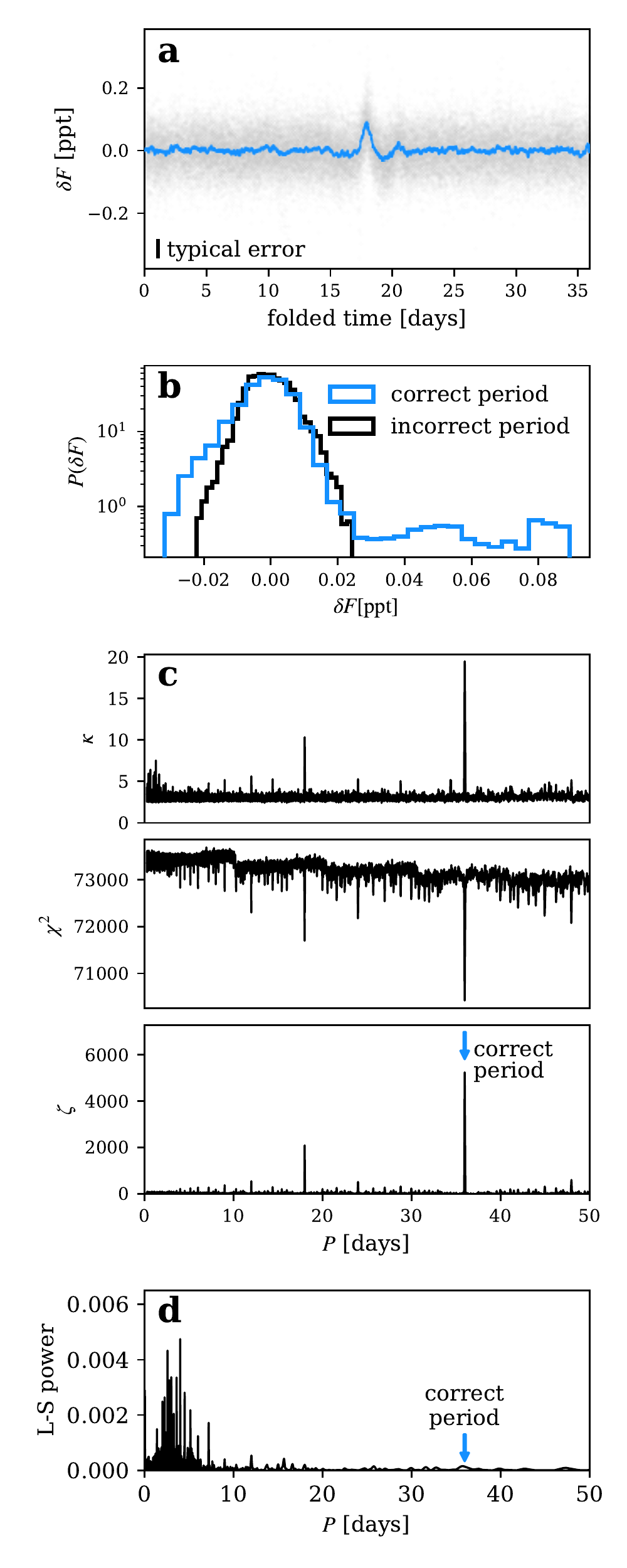}
    \caption{
	\textbf{a:} The phase curve of KIC 8694536 folded on $P = 35.95$ days with the
	candidate signal in blue.
    \textbf{b:} Histograms of the distribution of $f_i'$ when the light curve
	is folded on the correct ($P = 35.95$\,days) and incorrect ($P = 30$\,days)
	period.  Note the presence of heavy tails on the distribution from folding on
	the correct period.
    \textbf{c:} $\kappa$, $\chi^2$, and $\zeta$, which combines the two, as a
	function of period.
    \textbf{d:} \revision{The Lomb-Scargle periodogram for the same signal, which
    does not have significant power at the correct period.}
}
    \label{fig:example}
\end{figure}

For each period in our search grid, \sebastien\  first folds the data to construct a phase
curve. We then use a rolling mean over the fluxes (weighted by their
photometric errors) with a constant phase window of width $\Delta \phi$ to
construct a smoothed phase curve (hereafter a \emph{candidate signal};
e.g. Figure~\ref{fig:example}a). Specifically, for each point $i$ in a phase
curve with $N$ points, it computes the value of the candidate signal 
\begin{equation}
F'_i = \frac{\sum_{j=1}^N F_j w_{ij}}{\sum_{j=1}^N w_{ij}}
\end{equation}
where
\begin{equation}\label{eq:weights}
w_{ij} = \sigma_i^{-2} \Theta(\Delta \phi - |\phi_i - \phi_j|).
\end{equation}
Here $\phi_i$, $F_i$, and $\sigma_i$ are the phase, flux, and photometric
error, respectively, of the $i$th point and $\Theta$ is the Heaviside step
function.  For convienience, we define the relative flux as 
$\delta F = F/\bar{F} - 1$.

For each candidate signal, \sebastien\  computes two statistics: $\chi^2$, which reflects
how \emph{coherent} the phase curve is, and $\kappa$, which quantifies the
\emph{morphology} of the candidate signal. Specifically, 
\begin{equation}
\chi^2 = \sum_{i=1}^N \left( \frac{F_i - F'_i}{\sigma_i} \right)^2.
\end{equation}
Since the values of ${F'_i}$ are calculated directly from the values of $F_i$
themselves, this describes the scatter (dispersion) of flux values at a
constant phase. \revision{This is what our method has in common in with the family of 
PDM techniques.}  The other statistic, $\kappa$, is the the fourth standardized 
moment, or kurtosis, of the distribution of flux values in the candidate signal
(see Figure~\ref{fig:example}b). Specifically, 
\begin{equation}
\kappa = \frac{1}{N} \sum_{i=1}^N \left( \frac{F'_i - \mu}{\sigma} \right)^4
\end{equation}
where $\mu$ and $\sigma$ are the mean and standard deviation, respectively, of
the distribution of fluxes in the candidate signal. Kurtosis quantifies roughly
the ``tailedness" of a distribution; flux distributions of candidate signals
with low duty cycles will have a dominant mode around $\delta F = 0$, with
outliers on one or both sides caused by the signal.  \revision{Fig \ref{fig:example}d 
demonstrates that while both $\kappa$ and $\chi^2$ peak at the correct period,
a Lomb-Scargle periodogram does not pick out the signal, since it is not
well-described by a sinusoid.}

\begin{figure}
    \centering
    \includegraphics[width=0.46\textwidth]{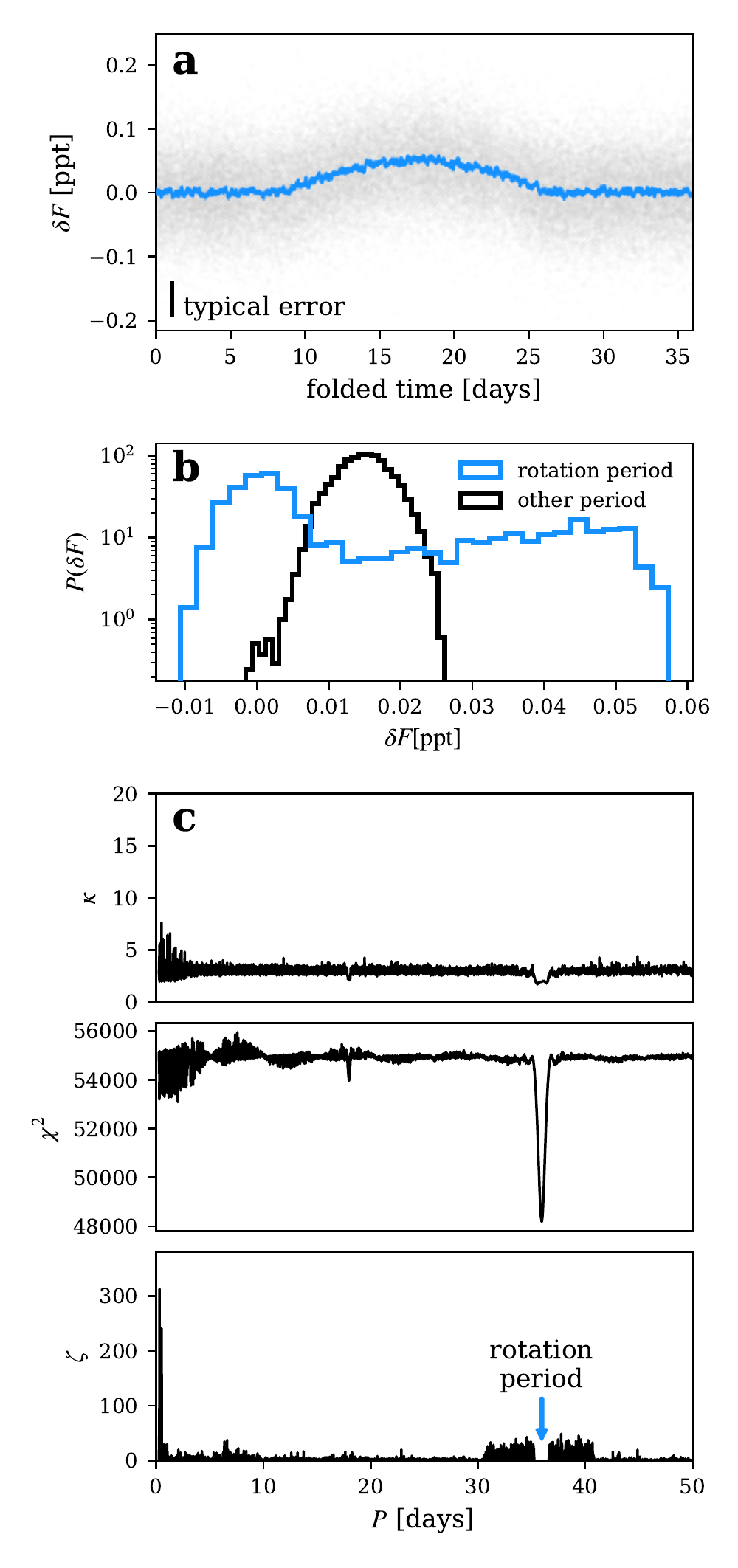}
    \caption{Same as Figure \ref{fig:example}, but for a simulated ideal rotator
        (i.e. a pure sine wave with all negative values set to 0).  Note
        that the \sebastien\ periodogram does not have a peak corresponding to 
        the injected rotation signal.
	\textbf{a:} The phase curve of the simulated data folded on $P = 35.95$ days with the
	candidate signal in blue.
    \textbf{b:} Histograms of the distribution of $f_i'$ when the light curve
	is folded on the correct ($P = 35.95$\,days) and incorrect ($P = 30$\,days)
	period.  
    \textbf{c:} $\kappa$, $\chi^2$, and $\zeta$, which combines the two, as a
	function of period.
}
    \label{fig:sine_example}
\end{figure}

The fourth standardized moment is the lowest that will work for our purposes,
since the third (skewness) is insensitive to $F'_i$ drawn from symmetrical
distributions, and the first and second standardized moments are identically
zero and one, respectively. The standard deviation of the fluxes in the
candidate signal (i.e. the \emph{unstandardized} centered second moment) will
pick out signals with large phase-integrated amplitude, but will not strongly
prefer short duty-cycle signals over those more extended in phase, \revision{such as the
flux modulations from rotating stars (see Figure \ref{fig:sine_example}a).} Conversely,
to the extent that the phases assigned to each point are random (i.e. the
period is wrong) the distribution of fluxes in the candidate signal will be
roughly Gaussian, since they are defined as the average of many randomly chosen
flux measurements.

From these quantities, we calculate $\kappa' \Delta \chi^2$ at every period,
where $\kappa' = \kappa-3$ is excess kurtosis, kurtosis less that of a Gaussian,
and $\Delta \chi^2$ is the local decrease in $\chi^2$ (see below for details).  
\revision{We chose to use the product, rather than a weighted sum of these
quantities for two reasons.  First, we wish to flag periods at which
\emph{both} statistics deviate from their baseline value. This is the same
reason we use $\kappa'$ rather than $\kappa$ and $\Delta \chi^2$
rather than $\chi^2$.  A null value of one statistic will ``cancel out'' a
non-null value of the other (see, for example, Figure \ref{fig:sine_example}c).
Second, a sum of $\Delta \chi^2$ and $\kappa'$
would require choosing relative weights, a tuning parameter that we prefer
to avoid.}
Calculating this
requires some model of the ``signal-free'' or ``baseline'' relationship between
$\chi^2$ and period. For uniformly spaced data (such as \kepler\ light
curves), we found that $\chi^2(P)$ takes discontinuous steps whenever the
duration of the smoothing kernel is a multiple of $\Delta t$, the light curve
cadence. This occurs whenever $P$ is a multiple of $\Delta t / \Delta \phi$.
For our application to \kepler, this is
\begin{equation}
\frac{\Delta t}{\Delta \phi} = \frac{0.02042\,\mathrm{d}}{0.002} = 10.208335\,\mathrm{d}.
\end{equation}
We fit a piecewise-linear function with discontinuities at multiples of
$\Delta t / \Delta \phi$ to obtain $\widehat{\chi^2}(P)$, then calculated
$\Delta \chi^2 (P) = \widehat{\chi^2}(P) - \chi^2(P)$. 
\revision{As mentioned later in section 2.4, this 
procedure does not correct perfectly for the discontinuous behaviour in $\chi^2$.
A better understanding of the mechanism causing these discontinuities would presumably 
suggest a better way to correct for or prevent them, but we were unable to 
determine their source and will leave its determination to future work.}

For each light curve, we construct a scrambled time series by randomly
re-assigning times to flux-uncertainty pairs. This preserves the distribution
of fluxes in the time series while erasing all patterns in the time domain.
We use \sebastien\  to calculate a periodogram from the scrambled data, from which we calculating
the moving standard deviation, which we use  to get the \emph{merit function}
(Figure~\ref{fig:example}c)
\begin{equation}
    \zeta(P) = \frac{\kappa'(P) \Delta \chi^2(P)}{\sigma(P)}
\end{equation}
where $\sigma(P)$ is the standard deviation of $\kappa'\Delta\chi^2$ values in
a small window surrounding P in the periodogram constructed from scrambled data
(we used a window of width 401 points centered around P, which spans roughly
$0.96 P$ to $1.04 P$). This quantity tracks, but is not equal to, the
signal-to-noise ratio (S/N) at each period (strictly S/N would be given by
the $\Delta\chi^2$ term alone). While the $\zeta$ might be interpreted
as a false-alarm probability (FAP) or p-value under certain assumptions, we caution
the reader against this. Real light curves have an abundance of periodic
structure from stellar rotation and variability, and the null hypothesis and
manifestly false in almost all cases. We calculate $\zeta$ not to obtain a FAP
for each detection, but to provide a way of ranking peaks from different
periodograms. An empirical investigation of the relationship between
false-positive rate and true-positive rate is presented in Section~\ref{sec:recovery}.
\revision{Note that $\zeta$ (as well as $\Delta \chi^2$ and $\kappa$ individually)
picks out the correct period, while a Lomb-Scargle periodogram (Figure 
\ref{fig:example}) does not.}

\subsection{Aliasing}

\begin{figure*}
\center
    \includegraphics[width=\textwidth]{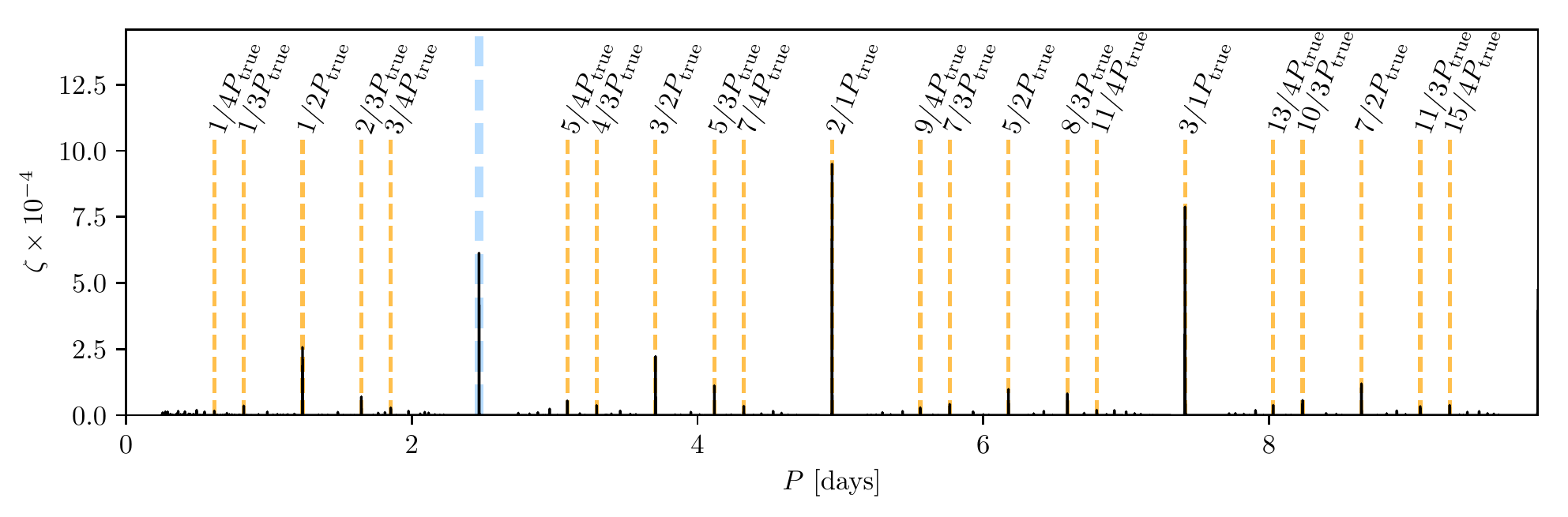}
    \caption{
	Periodogram generated from light curve of Kepler-1, also known are TrES-2
	\citep{odonovan:2006}, with aliases of Kepler-1b's period down to $n=3$
	labelled.  The correct orbital period is marked with the thick blue line.
    The presence of such aliases is why we consider only the highest peak from 
    each periodogram.
	}
    \label{fig:aliaslabels}
\end{figure*}

An inevitable outcome of using a folding technique with such a flexible model
is spurious peaks in the periodogram at rational fractions and multiples of the
``true'' period of a signal (Figure~\ref{fig:aliaslabels}). These peaks are
present in both $\chi^2$ and $\kappa$, since they come from the construction of
the candidate signal itself, rather than the operations performed upon it. We
see these peaks in both synthetic and real data.

It's worth noting that this is a different phenomenon from aliasing in the
context of the discrete Fourier transform, which is caused by finite sampling
of a continuous signal. It's also distinct from the phenomenon of harmonics
seen in the power spectra of non-sinusoidal signals.

\revision{
The height of each aliased peak decreases roughly with the 
denominator in the reduced rational form of it's period in units of 
$P_\mathrm{true}$, in other words it decreases with $n$, where
\begin{equation}
\frac{m}{n} = \frac{P_{\mathrm{flagged}}}{P_{\mathrm{true}}}
\end{equation}
is the reduced rational form of the ratio between flagged and true periods.
Examination of phase-folded light curves, shows that higher values of $n$ correspond 
to a lower-amplitude aliased signal and thus a smaller peak in the periodogram. }
This can be seen in
Figure~\ref{fig:aliaslabels}, which shows the periodogram for Kepler-1b with
aliases labelled. Because of the difficulty of automatically disambiguating
between aliases and true periods, we don't attempt to flag multiple signals per
light curve.

\subsection{Application to \kepler\ data}

We ran \sebastien\  on the instrumentals-removed PDCSAP fluxes on all
long-cadence light curves in \emph{Kepler} DR25 with at least twelve of 16
quarters present, not counting Q0 and Q17, which we don't use (161,786 stars).
We performed automatic outlier rejection by removing all points more than
$5 \sigma_{\mathrm{MAD}}$ away from the median within an 11 point window, where
$\sigma_{\mathrm{MAD}}$ is the standard deviation estimated by the mean
absolute deviation:
\begin{equation}
\sigma_{\mathrm{MAD}} = 1.4826 \times \mathrm{median}(|F_i - \mathrm{median}(\mathbf{F})|).
\end{equation}
We also removed all points with non-zero quality flags and split each time
series into segments at quarter boundaries and anywhere with a missing-data gap
of 0.3 days or greater. Within each segment, we used linear interpolation to
fill in any missing data, then detrended with a 2-day (roughly 99 cadences)
temporally-windowed rolling median to remove long duty-cycle signals and
trends. After detrending, we removed the interpolated points.

To compute periodograms for each star in our input catalog, we folded on a grid
of 26,492 period values from 0.25 days up to 50 days, with sampling uniform in
$\log P$ (motivated by experiments with synthetic data showing that periodogram
peak width increases thusly). Although we are primarily looking for signals
with periods in this range, \sebastien\  is still sensitive to signals with
shorter or longer periods if they have a strong alias in our period grid. We
constructed our candidate signals with $\Delta \phi = 0.002$ (see
Equation~\ref{eq:weights}), which corresponds to roughly 120 points in a
\kepler\ light curve with no missing data. 
\revision{The number of points in a constant-width phase bin of a phase curve
is not dependant on the folding period $P$, since folding amounts roughly to a 
reordering of points. }

\begin{figure}
    \centering
    \includegraphics[width=0.47\textwidth]{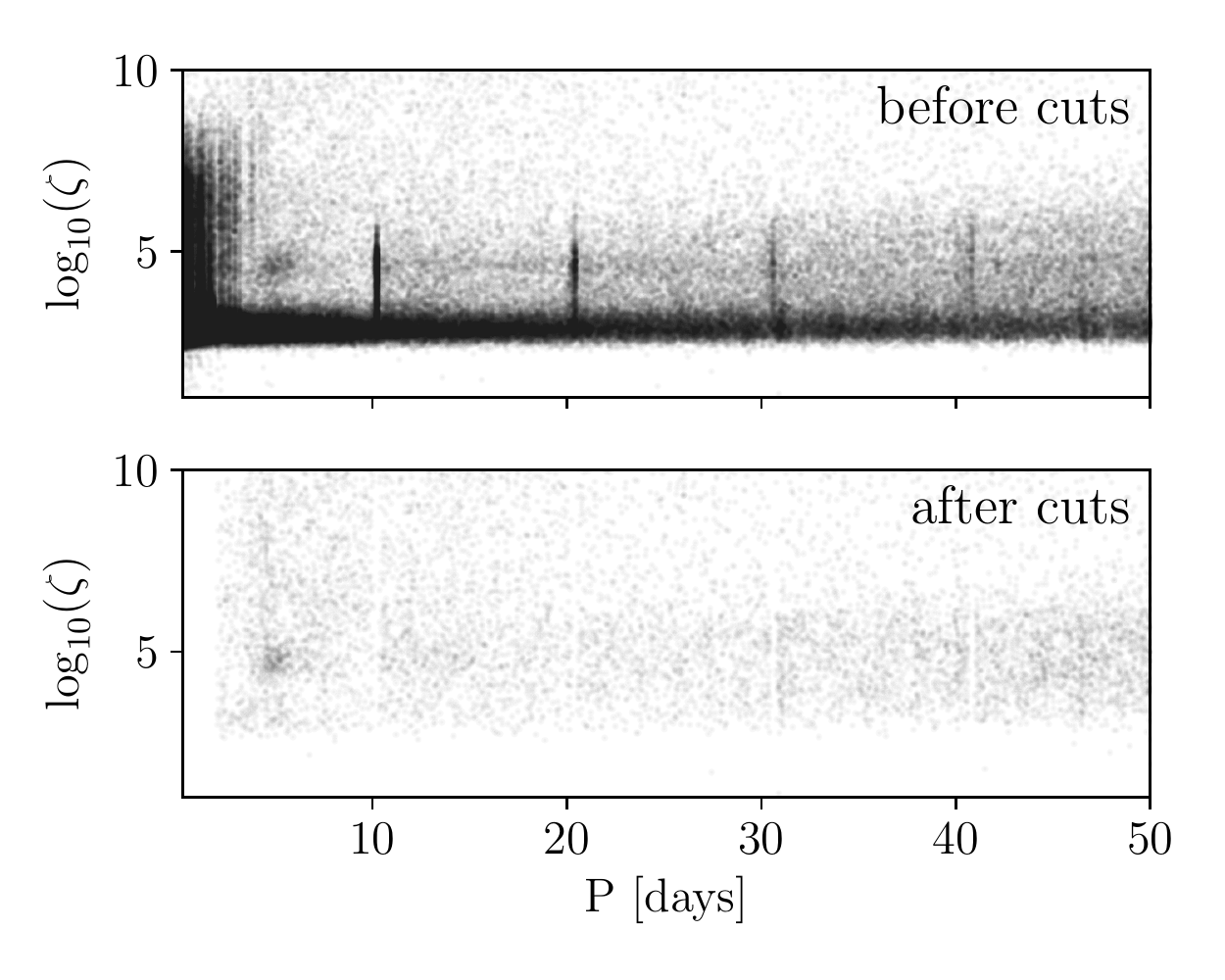}
    \caption{
	Flagged period values before and after making cuts. The
	overdensity of signals around $P \approx 5$ days is caused by our
	approximation of $\widehat{\chi^2}$. Manual examination of randomly selected
	events from this region revealed no unique characteristics.
	}
    \label{fig:peakscatter}
\end{figure}

We consider only the most significant peak from each light curve in order to
avoid the problem of disentangling aliases from independent signals. We make a
few cuts in order to remove as many spurious signals as possible (see
Figure~\ref{fig:peakscatter}). First, the 200 period values most frequently
flagged are eliminated on the grounds that they are likely an artifact of the
data analysis. These period values represent only 0.75\% of the period grid
over which we search. 
\revision{The number of removed periods was somewhat ad-hoc, but we manually 
    examined random signals from this population to ensure that they were 
    contaminated by spurious artifacts of the \kepler\ cadence.}
Second we remove any signals with $P < 2$\,days or with
$P$ too close to a discontinuity in $\widehat{\chi^2}$ (specifically,
$P / (1\,\mathrm{d})$ in any of (10, 10.3), (20.2, 20.5), (30.4, 30.7), or
(40.6, 40.9)).

We cross-matched the resulting catalog by \kepler\ input catalog number (KIC) with the Villanova eclipsing
binary catalog \citep{kirk:2016,abdul:2016}, \kepler\ DR25 KOIs and TCEs
\footnote{\revision{Some of the signals we flagged were TCEs in \kepler\ DR24, but were removed in DR25.  Additionally some are flagged in the \kepler\ 
inverted TCE list \citep{coughlin:2017}}}
\citep{thompson:2018}, the candidates from \citep{huang:2013}, the long-period
candidates from the wavelet-based search of \citet{dfm:2016} and planet hunters
\citep{wang:2015}, and the ultra-short period planets in \citet{sanchis:2014}.
We cross-match with both candidates and false positives from all catalogs.
Finally, we throw out any signal with $\kappa < 5$ on the grounds that they are
very likely to be long duty-cycle signals like stellar rotation.
After removing these KICs, we were left with with 4,426 flagged signals.

\label{sec:results}

\section{Recovery of other catalogs}
\label{sec:recovery}

\begin{figure}
\centering
\includegraphics[width=0.4\textwidth]{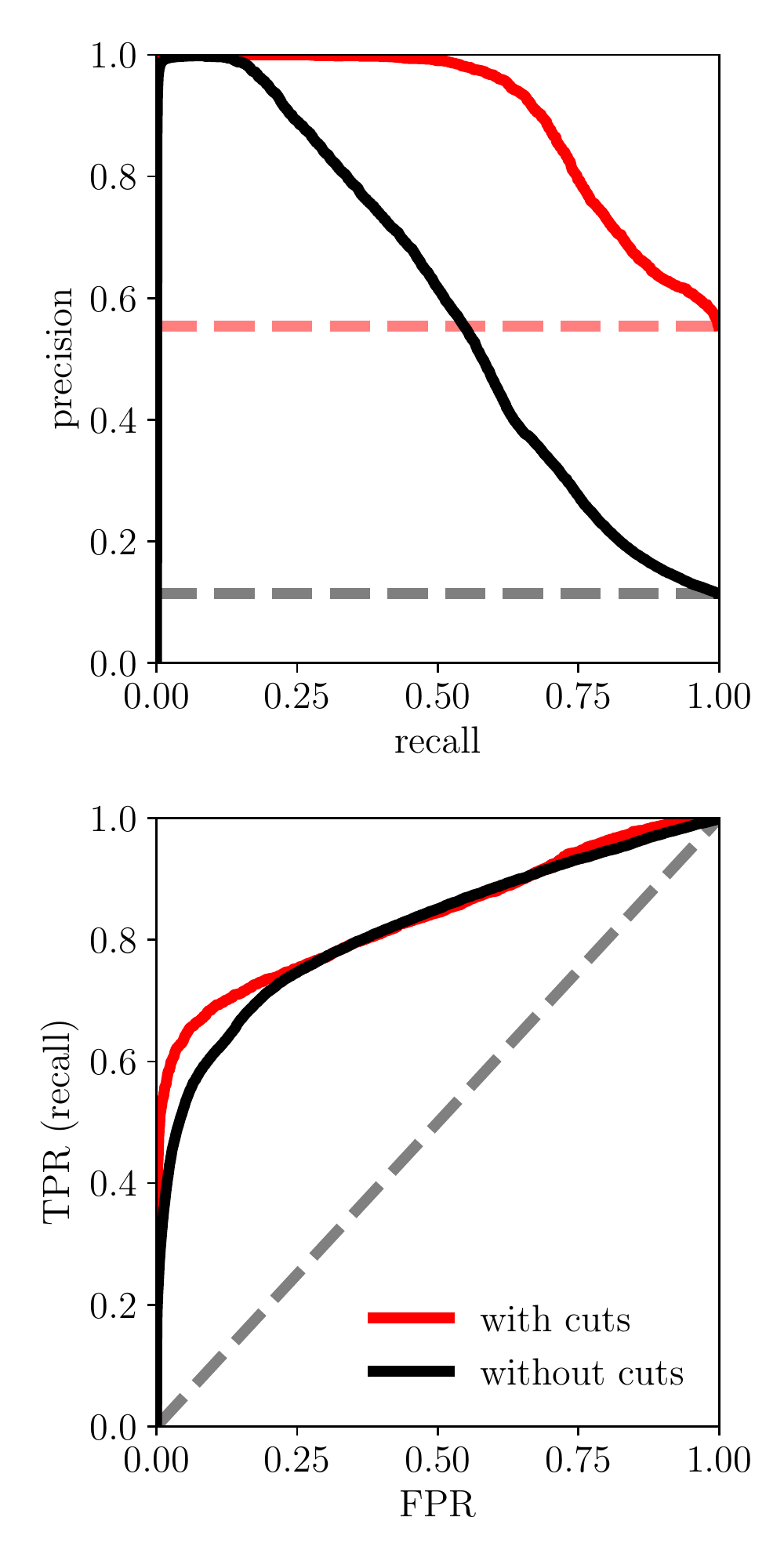}
\caption{
Precision-recall and receiver operating characteristic curves for our algorithm
applied to the \kepler\  data before and after making cuts.  \revision{In both plots, the
dashed lines show the expected performance of a classifier which flags light 
curves at random.}
}
\label{fig:recovery}
\end{figure}

As a way of quantifying the {\tt weirddetector}'s performance, we analyzed its
performance as a binary classifier of \kepler\ light curves. 
\revision{While injection-recovery
testing is commonly used to test new detection algorithms of various types, we
avoided this approach since our goal was not to recover a specific signal, and
there is no obvious signal type to inject.} Framing our
algorithm as a binary classifier relies on the fact that we attempt to flag at
most one signal per light curve. \revision{We consider a signal flagged if it's 
periodogram contains a peak with $\zeta > \zeta_\mathrm{threshold}$, and we
evaluate our recovery of signals in other catalogs as we vary 
$\zeta_\mathrm{threshold}$ }
A signal is considered an base-truth (expected) positive if it is in any of the 
catalogs with which we cross-matched
\citep{kirk:2016,abdul:2016,thompson:2018,huang:2013,sanchis:2014}, 
save \citet{wang:2015} and \citet{dfm:2016}, which include
nonperiodic signals. In this context, a false-positive is a distinct concept from a
scientific-false positive and has not been manually vetted.  Rather, it's 
a light curve with flagged by \sebastien\ with no detections in the cross-matched
catalogs.
Likewise, in this context, a true-positive can be a
\emph{scientific} false-positive if our algorithm flagged a signal that was also
flagged but listed as a scientific false-positive in a cross-matched catalog.
In other words, we are adopting the approximation that the set of interesting
signals (true positives) is exactly the union of the signals in the
cross-matched catalogs. Although the majority of these signals are transits or
transit candidates and thus better flagged by a model-based approach,
\sebastien\ would have recovered them equally well if they
were inverted or otherwise morphologically perturbed. 

Figure~\ref{fig:recovery} shows precision-recall and receiver operating
characteristic (ROC) curves for our algorithm as a classifier of \kepler\ light
curves. If we define $Y$ and $\widehat{Y}$ to encode the base-truth and
classifier output as
\begin{equation}
Y = \begin{cases}
    1 & \text{signal present} \\
    0 & \text{signal not present}
\end{cases}
\end{equation}
and 
\begin{equation}
    \widehat{Y} = 
    \begin{cases}
        1 & \text{light curve flagged} \\
        0 & \text{light curve not flagged}
    \end{cases},
\end{equation}
we can define precision, recall (true positive rate; TPR), and false positive
rate (FPR) as
\begin{equation}
    \text{precision} = P(Y = 1 | \widehat{Y} = 1),
\end{equation}
\begin{equation}
    \text{recall} = \text{TPR} = P(\widehat{Y} = 1 | Y = 1),
\end{equation}
and
\begin{equation}
    \text{FPR} = P(\widehat{Y} = 1 | Y = 0).
\end{equation}

The precision-recall and ROC curves show how these values change and the
classification threshold in $\zeta$ is changed. While precision-recall curves
are often more informative for needle-in-a-haystack problems where $P(Y = 1)$
is very low, their sensitivity to the number of base-truth positives
(interesting signals, i.e. $Y = 1$) in the sample can make them less
informative across data sets. The ROC curves quantify the performance of our
algorithm in a way that is independent of the fraction of interesting signals.
From both curves, it's clear that we are able to recover slightly over half of
the signals in the cross-matched catalogs with very minimal false-positive
contamination, \revision{which we can expect to be true for datasets with a similar
fraction of true positives.}

\section{Novel Detections}
\label{sec:detections}
After filtering these KICs out, most of what remained
were artifacts of stellar activity and rotation. We examined all the signals by
eye, and identified 52 that didn't appear to be the result of flares, rotation,
or otherwise spurious. Additionally, we manually determined the true period for
each signal since an alias had been flagged in many cases. For some signals,
the flagged periodic signal was induced by a short-duration nonperiodic one-off
event. Although it was not designed to recover such signals, \sebastien\  is
sensitive to them, since a high amplitude one-off event will still produce an
excursion away from the baseline flux in the candidate signal for a folded
phase curve.

We manually vetted each of these signals with three false positive tests (see
Section~\ref{sec:FPs}), yielding a total of 32 signals that survived as
candidates (hereafter weird objects of interest, or WOIs). Of the 33 WOIs, 18
are periodic signals that appear to be caused by neither transiting bodies nor
stellar rotation (Section~\ref{sec:HBs}), \lrevision{6} are single or double transit-like
events (Section~\ref{sec:transits}), and 5 are ``brightening'' events
resembling inverted transits (Section~\ref{sec:IDs}). The remaining 4 WOIs
(including Boyajian's star; \citet{boyajian:2016}) are already reported in the
literature (Section~\ref{sec:knowns}).  \revision{While many of these signals 
    do not satisfy our periodicity assumption (the single dips and brightening 
    events, we have chosen to report them because similar events are likely to 
be flagged when applying this method to other data sets.}

\subsection{Periodic signals}
\label{sec:HBs}

\begin{figure*}
\centering
\includegraphics[width=0.95\textwidth]{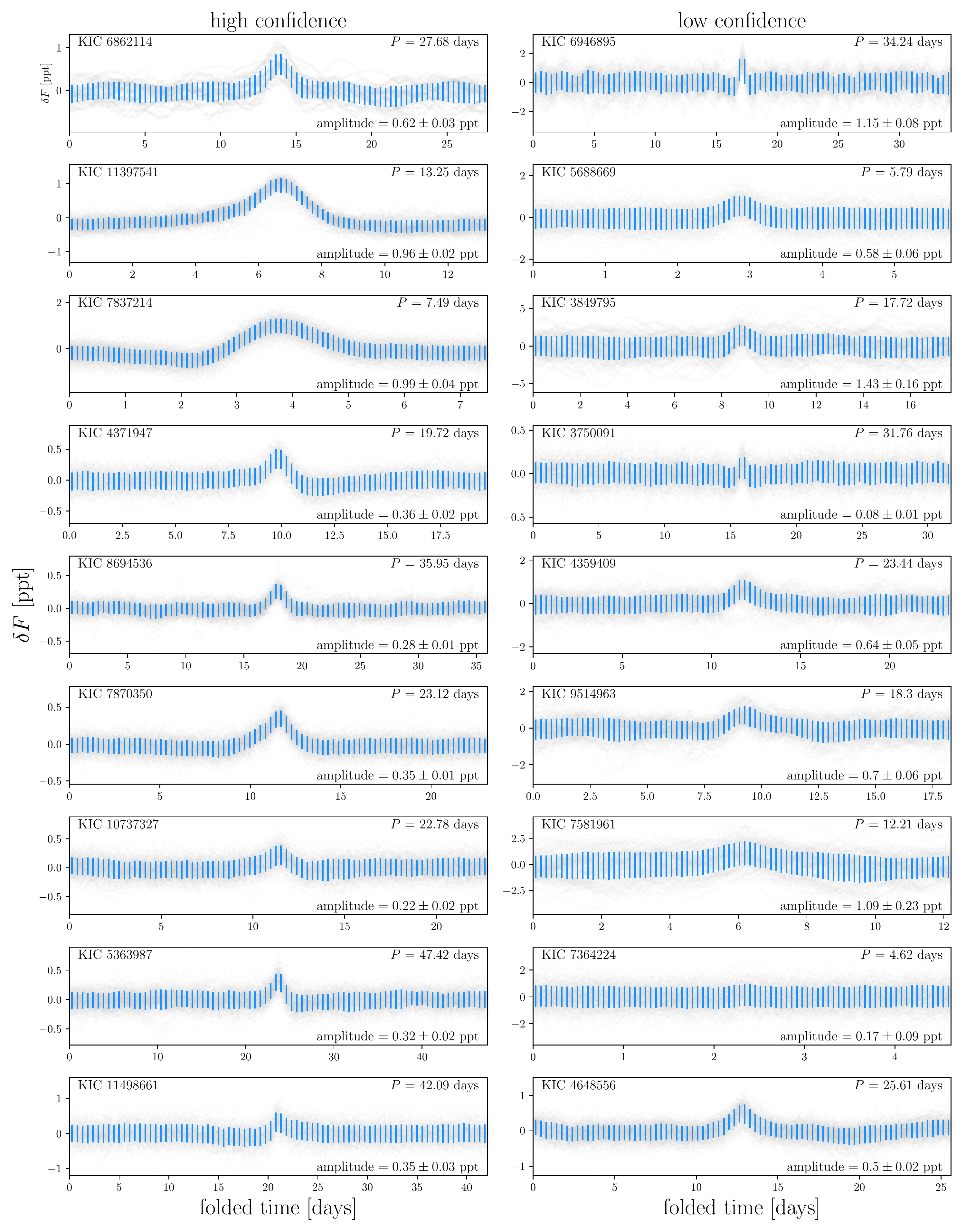}
\caption{
    Eighteen previously unknown periodic \kepler\ signals.  \textbf{left:} the high-confidence detections, \textbf{right:} the low-confidence detections, for which the timescale of stellar variability is roughly that of the flagged signal. The vertical blue lines show the mean and scatter of flux values in 80 constant-width bins in phase. The light curves have not had long-term trends removed.
}
\label{fig:HB}
\end{figure*}

\begin{table}
\caption{
\emph{Candidate signals that may be caused by heartbeat tides.  
\hspace{\textwidth} 
${}^a$flagged period: 45.5543(91)}}
\centering 
\begin{tabular}{l l l} 
\hline\hline 
KIC & $P$ [days] & $\zeta$ \\ [0.5ex] 
\hline 
        \multicolumn{3}{@{}l}{\emph{High confidence}}\\
        \hline  
        4371947 & 19.7175(39) & 926.17 \\
        5363987 & 47.4229(95) & 94.25 \\
        6862114 & 27.6799(55) & 5197.9 \\
        7837214 & 7.4895(15) & 1221.98 \\
        7870350 & 23.1156(46) & 226.8 \\
        8694536 & 35.9492(72) & 312.45 \\
        10737327 & 22.7772(46)${}^a$ & 115.59 \\
        11397541 & 13.2541(27) & 2572.16 \\
        11498661 & 42.0856(84) & 27.08 \\
        \hline
        \multicolumn{3}{@{}l}{\emph{Low confidence}}\\
        \hline
        3750091 & 31.7631(64) & 352.57 \\
        3849795 & 17.7238(35) & 2864.48 \\
        4359409 & 23.4415(47) & 190.97 \\
        4648556 & 25.6132(51) & 109.8 \\
        5688669 & 5.7852(12) & 6100.39 \\
        6946895 & 34.2370(68) & 28801.39 \\
        7364224 & 4.62144(92) & 136.02 \\
        7581961 & 12.2131(24) & 155.09 \\
        9514963 & 18.3001(37) & 166.52 \\ [1ex]
\hline\hline 
\end{tabular}%
\label{tab:heartbeats} 
\end{table}

\begin{figure*}
\centering
\includegraphics[width=0.95\textwidth]{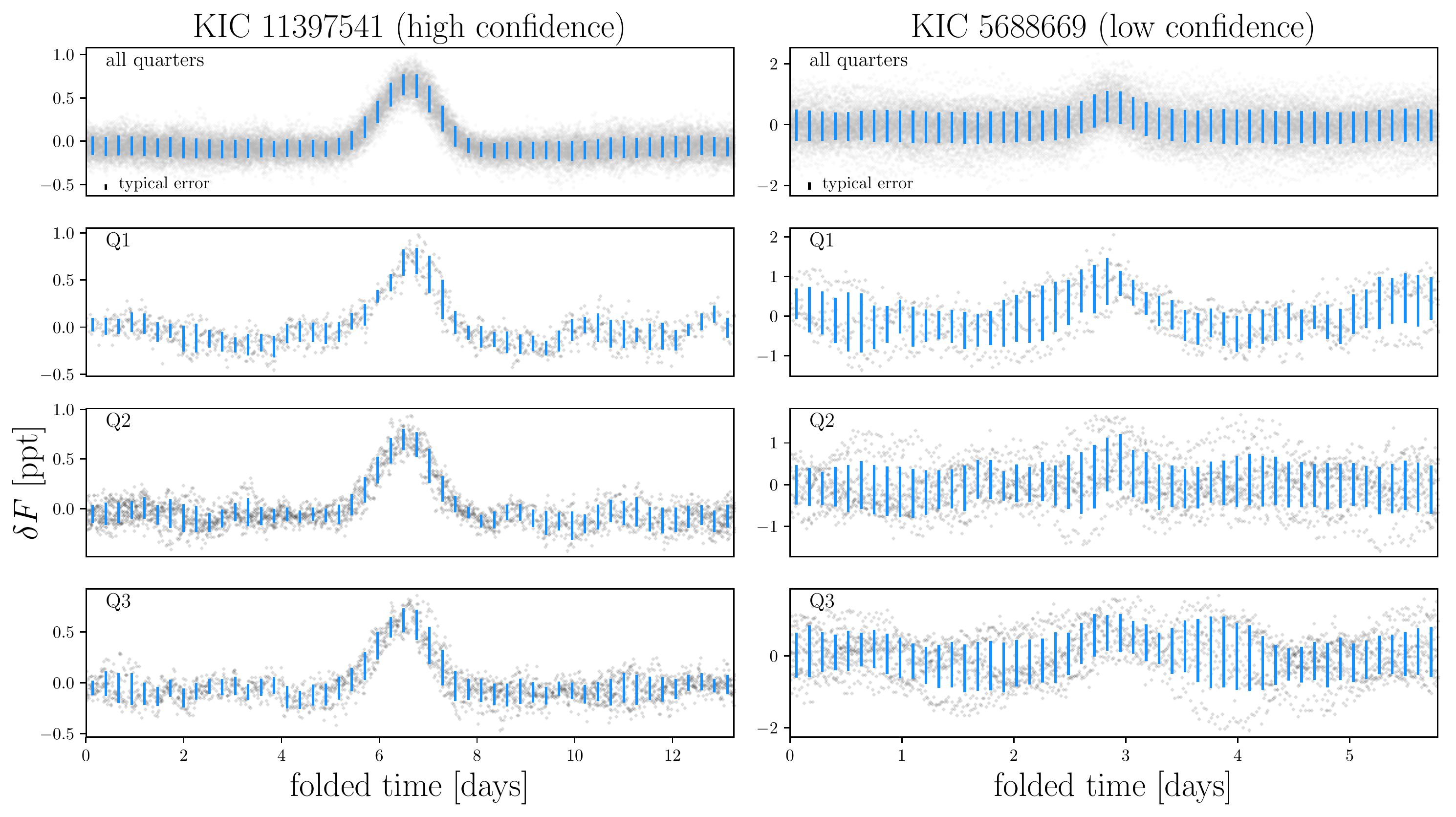}
\caption{
    \revision{An example of a high-confidence (KIC 11397541) and low-confidence (KIC 11397541)
    periodic candidate signals, along with phase-folded light curves from quarters
    1-4.  While both signals appear to be real when folded on all \kepler\ data,
    The phase-folded light curves for KIC 11397541 reveal that this signal is not 
clearly present in the data, but may be an artifact of the finite \kepler\ baseline.}
}
\label{fig:confidence}
\end{figure*}

We flagged 18 periodic signals with $P<50$\,days, which we divided into
\emph{high-confidence} and \emph{low-confidence} groups of 9 signals each
(see Fig.~\ref{fig:HB}, Table~\ref{tab:HB}). The low-confidence group is not distinguished by
low $\zeta$ values, \revision{but by the presence of stochastic variability with 
timescales comparable to those of their candidate periodic signals (see 
Figure~\ref{fig:confidence} for contrasting examples of each).  Given the finite
\kepler\ baseline, we do not want to rule out the possibility that signals such as 
those in  our low-confidence group could spuriously arise.}

We hypothesize that a plausible interpretation for many of the WOI periodic
signals is that they are systems exhibiting heartbeat tides and in
non-eclipsing configurations. Heartbeat binaries (which are named for the fact
that their light curves often resemble electrocardiograms) are a class of
highly eccentric binaries that display growing and shrinking ellipsoidal
variations over the course of an orbit \mbox{\citep{welsh:2011,thompson:2012}}. The
light curves of heartbeat binaries display a variety of temporal
morphology not present in our flagged sample, but it's possible that this is
caused by a detection bias of our algorithm or of the cross-matched catalogs.
For example a heartbeat binary that exhibits a periodic decrease in flux,
rather than an increase, is much more likely to be a \kepler\ threshold
crossing event (TCE), and thus thrown out by us.

\citet{penoyre:2018} recently highlighted the possibility that heartbeat tides
could be caused by highly eccentric massive planets, and our smaller signals
($\delta F \approx 5\times 10^-4$) are in the regime that they calculate could
be caused this way. We emphasize, however, that if these signals are indeed
caused by heartbeat tides, stellar companions are generally more likely to be
the culprit. \revision{In order to calculate the amplitude of each signal (reported 
    in Figure \ref{fig:HB}), we fit it with a sum of 
harmonic oscillations with periods ${P, \frac{P}{2}, \ldots , \frac{P}{30}}$, where
$P$ is the period of the signal.}

These signals are coherent over the 1344 days of \kepler's first sixteen
quarters, a feature which strongly suggests they aren't induced by the
combination of starspots and stellar rotation \citep{giles:2017}. Furthermore,
they generally have a very short duty-cycle, e.g. they are ``on'' in a small
fraction of total phase---a feature that which is unlikely to be produced by
rotation with persistent bright or dark surface features.

\subsection{Single and double transit candidates}
\label{sec:transits}

\begin{figure}
\centering
\includegraphics[width=0.49\textwidth]{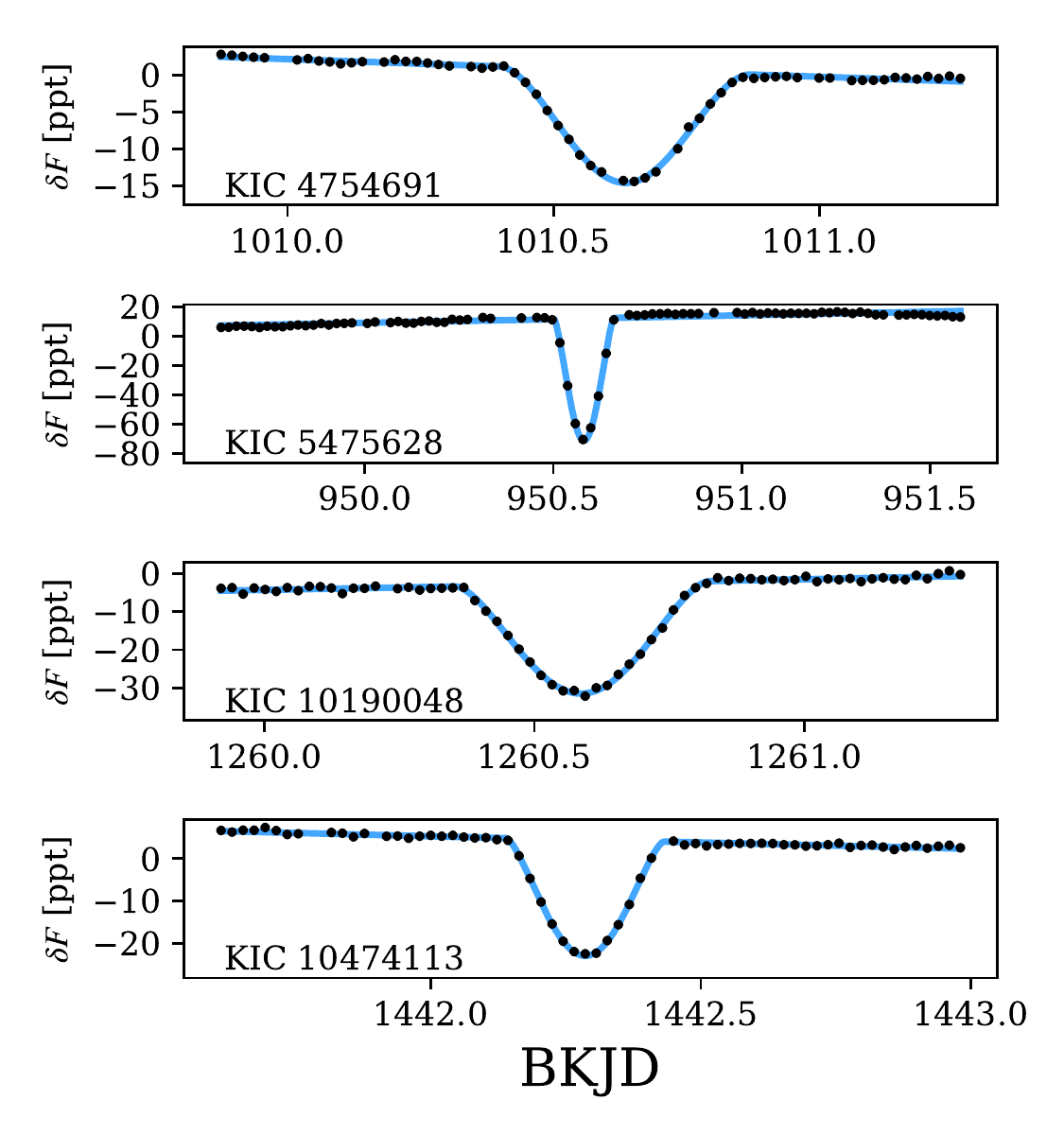}
\caption{
    Single transit events, with maximum a posteriori (MAP) fit in blue. Photometric
errors are accounted for but too small to be visible.
}
\label{fig:SD}
\end{figure}

\begin{figure*}
\centering
\includegraphics[width=0.9\textwidth]{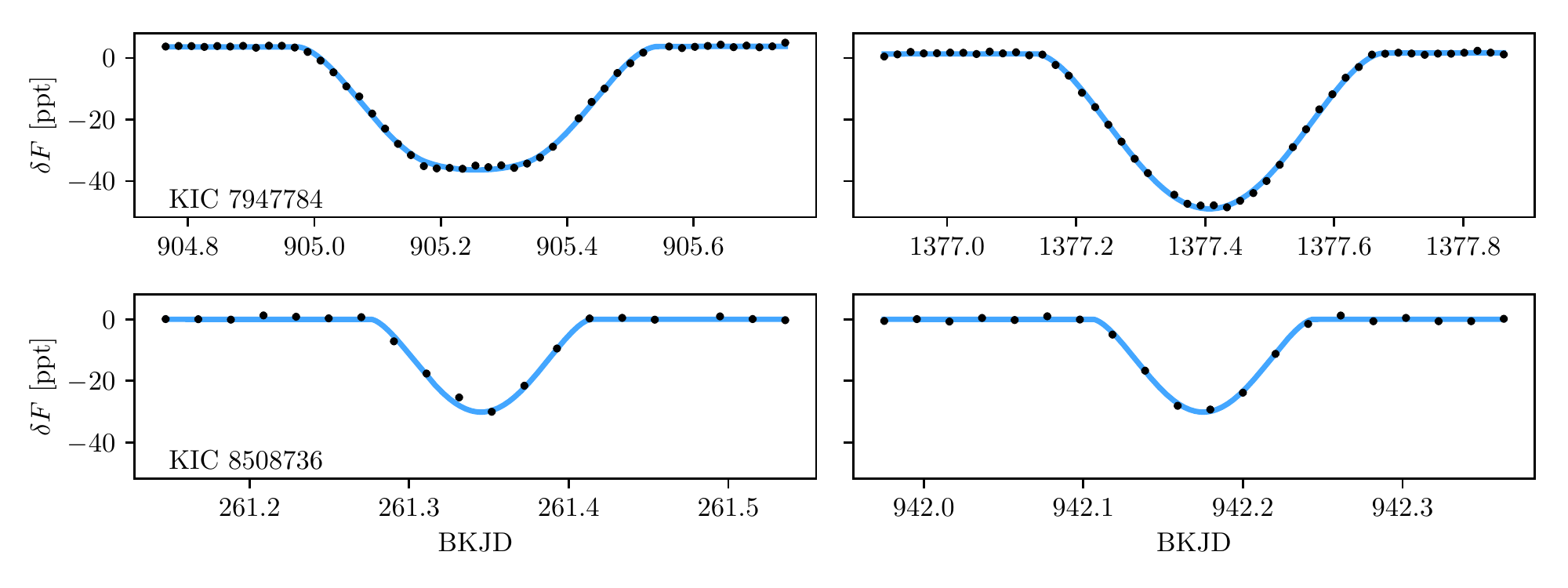}
\caption{
Double transit events, with maximum a posteriori fit in blue. KIC 7947784 was
fit with two singly-transiting planets, while KIC 8508736 was fit with a single
planet. Photometric errors are accounted for but too small to be visible.
}
\label{fig:DD}
\end{figure*}

\begin{table*}
\caption{
\emph{\lrevision{Six} new single and double transiting system candidates. Note that KIC 7947784 displays two transits but is actually likely two distinct orbiting bodies.} 
} 
\centering 
\begin{tabular}{l l l l l} 
\hline\hline 
KIC & $\zeta$ & {epoch [BKJD]} & $P$ [days] & $R_P$ $[R_{\odot}]$ \\ [0.5ex]  
\hline 
        \multicolumn{5}{@{}l}{\emph{Single transits}}\\
        \hline
        4754691  & 307.38 & $1010.63348^{+0.00036}_{0.00036}$  & $2875^{+757}_{-557}$ & $\geq 0.206^{+0.082}_{-0.075}$\\
        5475628  & 89.07 & $950.58131_{-0.00035}^{+0.00032}$ & $975^{+309}_{-221}$ & $\geq 0.536^{+0.109}_{-0.114}$ \\
        10190048 & 634.99 & $1260.58772^{+0.00083}_{-0.00077}$ & $1508^{+325}_{-215}$ & $\geq 0.262^{+0.092}_{-0.093}$\\
        10474113 & 433.29 & $1442.857365^{+0.00039}_{-0.00036}$ & $2392^{+2763}_{-828}$ & $\geq 0.568^{+0.278}_{-0.212}$\\
        \hline
        \multicolumn{5}{@{}l}{\emph{Double transits}}\\
        \hline
        7947784 & 1303.57 & $905.25744^{+0.00046}_{-0.00045}$ & $1451^{+53}_{-42}$ & $ \geq 0.223^{+0.051}_{-0.040}$ \\
        		& 1303.57 & $1377.40631^{+0.00044}_{-0.00044}$ & $1270^{+38}_{-18}$ & $ \geq 0.399^{+0.120}_{-0.088}$ \\
        8508736 & 1078.21 & $261.34493^{+0.00044}_{-0.00046}$ & $680.83003_{-0.00062}^{+0.00062}$ & $ \geq 0.382^{+0.139}_{-0.141}$\\
\hline\hline 
\end{tabular}
\label{tab:singles} 
\end{table*}

To characterize our \lrevision{four} single and two double-transit WOIs, we used all data
within 0.7 days of the transit midpoint, with no detrending. We generated transit
models with \batman\ \citep{kreidberg:2015}, and performed the inference with
\multi\ \citep{feroz:2008,feroz:2009,feroz:2013}. All transits were fit jointly
with a linear baseline to account for long-term trends. We sampled
from quadratic limb-darkening laws using the uninformative parametrization from
\cite{kipping:2013}. We calculate the log-likelihood by treating measured
fluxes as drawn from independent heteroscedastic normal distributions centered
on their true values. Specifically,
\begin{equation}
\log(\mathcal{L}(\theta)) = - \sum_i \frac{1}{2} \left ( \frac{F_i - \mu_i(\theta)}{\sigma_i} \right)^2 + C,
\end{equation}
where $f_i$ and $\sigma_i$ are the flux and photometric error of the
$i^{\mathrm{th}}$ point, $\mu_i(\theta)$ is the flux value at the phase of the
$i^{\mathrm{th}}$ point as calculated by \batman\ for the parameters $\theta$,
and $C$ is a constant independent of $\theta$. Our \multi\ output, including
posterior samples, \revision{as well as detailed prior documentation}, is available online 
at \datalink.

\lrevision{All four single dips} favor grazing transit geometries, and display
excellent fits as seen in Figure~\ref{fig:SD}. Consequentially, their impact
parameters, $b$, are nearly unconstrained from above and are in practice
limited by the arbitrary maximum value set by our prior ($b_{\mathrm{max}} =
1.5 R_{\star}$). This truncation of parameter space effects posterior
estimation of $R_P/R_{\star}$, which is strongly positively covariant with $b$,
and means that we can only provide lower bounds on planetary radii. To derive
these bounds, we combine samples from $p(R_P/R_{\star}|\mathrm{data})$ for each
orbiting body with the value of $R_{\star}$ reported in \cite{mathur:2017},
which is reported as a single value with upper and lower error bars. Given
$R_{\star} = \mu^{+\sigma_+}_{-\sigma_-}$, we adopt a density of
\begin{equation}
\label{eq:skeweddist}
p(R_* | \mu^{+\sigma_+}_{-\sigma_-}) = 
\begin{cases} 
    2 \frac{\sigma_-}{\sigma_- + \sigma_+} \mathcal{N}(R_*|\mu, \sigma_-) & R_* \leq \mu  \\
    2 \frac{\sigma_+}{\sigma_- + \sigma_+} \mathcal{N}(R_*|\mu, \sigma_+) & R_* > \mu
\end{cases}.
\end{equation}
This is equivalent to equation (3) of \cite{yi:2018}, save that we are not formally excluding negative stellar radii.

Inferred properties of singly-transiting bodies are typically very uncertain
because of the lack of constraining power on the orbital period. We adopt a
uniform prior in $\log P$, and a skewed-normal prior on $\rho$ taken from
\cite{mathur:2017} with a density of the same form as that given in
Equation~(\ref{eq:skeweddist}). The minimum orbital period for the prior was
set by using the \kepler\ light curve itself to check what periods were
excluded.  Our prior over all parameters not listed here is uniform over a 
finite window \revision{(see the online documentation for details)}.

We first ran our inference code on all single dips with eccentricity fixed
to zero and found that KIC 4754691 and KIC 10190048 were
well-described by this model (see Table~\ref{tab:singles} for the inferred
parameters). For KIC 5475628 and KIC 10474113, we find eccentricity is required
to explain the shapes, so we allowed eccentricity to vary with uniform
priors on $\sqrt{e}\sin(\omega)$ and $\cos(\omega)$ ($\omega$ being the
longitude of periastron).

KIC 8508736 has two transit-like events that are well-fit by a single orbiting
body. Because we observe the putative body at two epochs, its orbital period
is well-constrained to high precision. The other two transit-like events, in
the light curve of KIC 7947784, do not appear to share the same morphology -
as evident from the top panel of Figure~\ref{fig:DD}. It is possible that we
are seeing two transits of a single body exhibiting extreme precession.
However, in this scenario the period would be $472$\,days and thus should have
repeated in quarter 5 (yet clearly does not), unless the precession was so
exteme it evades transit for that epoch.
We consider it more likely we are witnessing two independent single transits
instead and model it as such here.

\subsection{Brightening events}
\label{sec:IDs}

\begin{figure}
\centering
\includegraphics[width=0.5\textwidth]{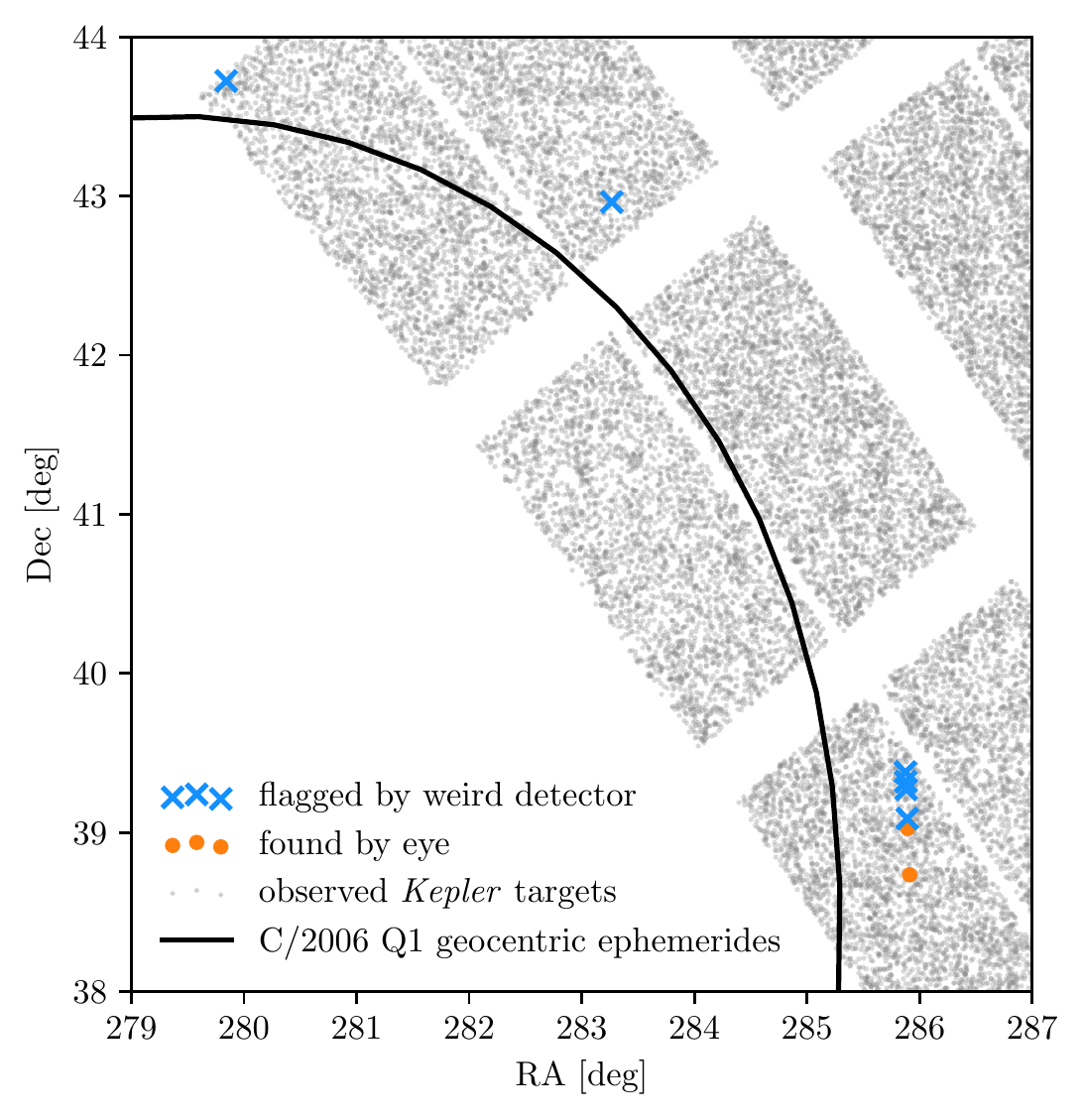}
\caption{
The path of C/2006 Q1 on the sky as viewed by a geocentric observer from HORIZONS, along with the ephemerides of our WOIs and false positives induced by the comet. \kepler's earth-trailing orbit means that it's ephemerides for solar-system objects are offset from geocentric ephemerides.
}
\label{fig:comet}
\end{figure}

\begin{figure}
\centering
\includegraphics[width=0.5\textwidth]{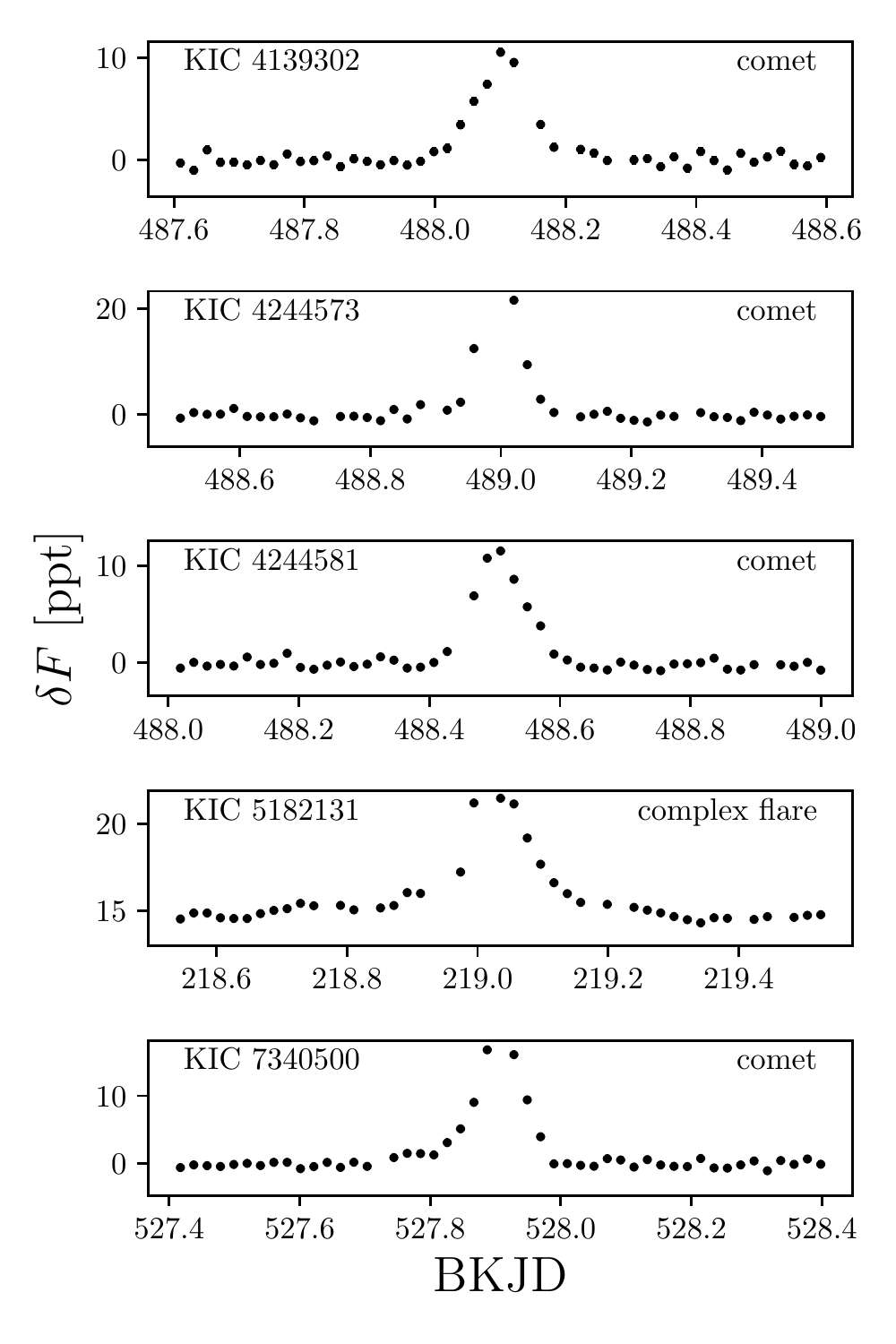}
\caption{Brightening events.  Photometric errors are accounted for but too small to be visible.  Four out of five brightening events are caused by passing solar-system comet C/2006 Q1, and one is likely a complex flare. }
\label{fig:brightenings}
\end{figure}

In total, we detected five KICs displaying brightening events, or
inverted transits. However, we find that four of these five WOIs
(Figure~\ref{fig:brightenings}) that passed our false positive tests
(KIC 4139302, KIC 4244573, KIC 4244581, and KIC 7340500) are caused by
the Solar System comet C/2006 Q1 passing through the
\kepler\ field-of-view. Two signals that were flagged as blends (from KIC
3937417 and KIC 7935479) are also caused by the comet.

The cometary nature of these events is easily recognizable because some of
their epochs are separated by a timescale of order days, and they trace out
a clear path on the sky. We searched for Solar System bodies by the event
ephemeris for one of our signals using JPL's online HORIZONS
tool\footnote{\url{https://ssd.jpl.nasa.gov/?ephemerides}}, from which
we obtained detailed ephemerides of the comet's path (see
Figure~\ref{fig:comet}). Because \kepler\ is on an Earth-trailing orbit
\citep{vancleve:2016}, its ephemerides for Solar System objects are different
from those of a ground-based observer.  This accounts for the systematic offset
between out events and the geocentric C/2006 Q1 ephemerides.

The presence of this comet in \kepler\ was reported in \cite{greist:2014},
along with that of C/2007 Q3 and 3 possible new comets, although none of the
light curves flagged by our algorithm were among those they reported. A
non-exhaustive manual search along the path of the comet yielded three more
contaminated light curves not reported in \cite{greist:2014}: KIC 3628785,
KIC 3937430, and KIC 3937432. We attribute the fact that we didn't recover
the KICs they flagged to our method being designed for the detection of
periodic signals, as well as the limitations of manual analysis of flagged
events.

The remaining brightening WOI, KIC 5182131, is not associated with C/2007 Q3
and we argue is likely caused by flaring.
Flares generally have rise times of a few minutes or less, and have a strongly
asymmetrical temporal morphology. Complex flares, which are made up of multiple
simultaneous flares, are the exception to this rule and are typically seen only
on particularly active stars \citep{davenport:2014}. KIC 5182131 exhibits frequent
classical flares, in contrast to the other stars with brightening events, so we
hypothesize that its event is a complex flare (see Figure~\ref{fig:brightenings}).
This event is not fit well by a
self-lensing binary model, (e.g. \citealt{kruse:2014}), which is equivalent to an 
inverted 
transit model when the Einstein radius of the lens is small \citep{agol:2003}.

\subsection{Previously reported signals}
\label{sec:knowns}


Besides those that are in the catalogs with which we cross-matched, we found
four signals that have been previously reported in the literature. We performed
the same false-positive tests on these as for the previously unreported
defections. Particularly notable is that fact that we recover KIC 8462852, also
known as Boyajian's star, since it was first detected with the Planet Hunters
crowd-sourced manual search \citep{boyajian:2016}.
KIC 10402660, a potential ultra-short period planet first reported in \cite{jackson:2013}, was flagged, as well 
as KIC 5793963 and KIC 11918466, which failed our false positive tests, but are reported as transit candidates in \cite{tenenbaum:2012}.

\subsection{False positives}
\label{sec:FPs}

\begin{table*}
\caption{
\emph{
Flagged signals that failed our false positive tests.  The \emph{reason} column states which false-positive test the signal failed (QAM: signal amplitude mismatch between quarters, EM: ephemeris match).
}} 
\centering 
\begin{tabular}{l l c l l l c} 
\hline\hline 
KIC & $\zeta$ & reason & $P_{\mathrm{true}}$ [days] & $P_{\mathrm{flagged}}$ [days] & epoch(s) [BKJD] & category/explanation\\ [0.5ex] 
\hline 
        2578184 & 12.17 & blend &  &  & 171.6 & brightening event \\
        2993038 & 572.64 & blend &  &  & 551.0, 1256.0 & single/double transit \\
        3937417 & 317.79 & blend &  &  & 486.5 & comet \\
        4143192 & 290.83 & EM: KIC 3836439 & 1.54035(31) & 6.1614(12) &  & short period planet \\
        4488363 & 1042.88 & blend &  &  & 1057.0, 1433.0, 766.0 & brightening event \\
        5120608 & 677.81 & blend &  &  & 1441.0 & brightening event \\
        5194255 & 167.66 & QAM & 18.1543(36) & 18.1543(36) &  & nontransit periodic \\
        5462390 & 1015.34 & blend &  &  &  & instrumental \\
        5793963 & 2471.58 & EM: KIC 6953219 & 0.56681(11) & 4.53446(91) &  & short period planet \\
        6044768 & 2029.32 & EM: KIC 6953219 & 0.56680(11) & 7.9352(16) &  & short period planet \\
        6125804 & 3346.49 & EM: KIC 6953219 & 0.56680(11) & 8.5020(17) &  & short period planet \\
        6525462 & 3001.53 & EM: KIC 6367628 & 3.77979(76) & 18.8989(38) &  & eclipsing binary \\
        6606591 & 3409.47 & EM: KIC 6953219 & 0.56680(11) & 8.5020(17) &  & short period planet \\
        6606797 & 943.74 & EM: KIC 6953219 & 0.56681(11) & 2.83405(57) &  & short period planet \\
        7867378 & 30.68 & blend &  &  & 912.0 & brightening event \\
        7935479 & 35.98 & blend &  &  & 555.5 & comet \\
        8114216 & 274.7 & QAM & 11.7977(24) & 11.7977(24) &  & nontransit periodic \\
        8694723 & 217.06 & crosstalk &  &  & 1280.4 & single/double transit \\
        9955874 & 187.74 & pixel sensitivity dropout & & & 895.5 & single dip \\
        11753409 & 486.83 & EM: KIC 11235323 & 19.6702(39) & 19.6702(39) &  & eclipsing binary \\
        11918466 & 1486.27 & blend & 8.0764(16) & 48.4584(97) &  & short period planet \\ [1ex]
\hline\hline 
\end{tabular}
\label{tab:FPs} 
\end{table*}

To vet the 52 manually picked signals, we performed three tests, yielding 20
false positives (see Table~\ref{tab:FPs}). First, we checked if the signal
matched the morphology and ephmerides of known KOIs or eclipsing
binaries in the Villanova eclipsing binary catalog (similar to the strategy
used by \citealt{coughlin:2014}). Those with a match we
presumed to be either a blend of the star with the known KOI or eclipsing
binary, or contaminated by video crosstalk (electrical interference).

In the case that no match was found, we manually examined their target pixel
data for signs of blending or crosstalk, which are evident when the signal is
not coming primary from the center of the aperture, but is offset or present in
only a few pixels. If the signal appears inverted in some pixels, we also take
this as an indication of crosstalk. Finally, for periodic signals, we also
examined the difference in signal amplitude between quarters and take strongly
varied amplitudes as indication of contamination.

\section{Discussion}
\label{sec:discussion}

Our work has introduced a new approach for flagging unusual astrophysical
signals in time series data, under the assumption of strict periodicity.
We exploit the fact that a strictly periodic signal is coherent with itself
when folded upon the correct period, allowing for its recovery via phase
dispersion minimization. As a demonstration, we have applied our algorithm
(\sebastien) to 161,786 \kepler\ light curves and uncovered 18 previously
unnoticed periodic signals, and several transiting system candidates, as
well as some ``contaminant'' signals from a passing solar-system comet and
stellar activity.

The search for unexpected behavior in data is presently not completely
automated, since interpretting such signals requires human judgement.
Nevertheless, the statistical properties of threshold crossing events
(i.e. candidate signals) can be rigorously determined with an automated
approach such as ours. For highly irregular signals, which fall within
the extremely-broad class of signals to which it is sensitive, our
technique is therefore expected to expedite the discovery process by
orders of magnitude.

As mentioned though, a current bottleneck in the discovery process of
irregular signals is the need for human interpretation of candidates
found by \sebastien. We highlight here that it may be possibe to treat 
\sebastien\  as a way of feature-engineering a light curve or set of light
curves, to produce a currated set of phase curves that serve as the input to
a machine leaning algorithm.

The vast majority of the by-eye analysis performed for this paper was to
categorize signals as either aligned flares or stellar rotation and to correct
the periods of any signals that were flagged as aliases. A machine-learning
approach to phase-curve classification could make it possible to construct a
nearly-fully-automated pipeline, making this approach more reproducible and
even more scalable to large datasets. Such an approach would effectively replace
the manual classification of phase curves in this work with an automated system,
much like how planetary candidates are often vetted automatically
(e.g. \citealt{thompson:2018}).

\section*{Acknowledgments}

DMK is supported by the Alfred P. Sloan Foundation. Thanks to the
Cool Worlds Lab members and for helpful discussions throughout the
course of this research.  Thanks to our anonymous reviewer for their 
helpful comments, from the which the paper greatly benefited.

We acknowledge computing resources from Columbia University's Shared Research Computing Facility project, which is supported by NIH Research Facility Improvement Grant 1G20RR030893-01, and associated funds from the New York State Empire State Development, Division of Science Technology and Innovation (NYSTAR) Contract C090171, both awarded April 15, 2010.

\bibliographystyle{mnras}
\bibliography{references}
\label{lastpage}
\end{document}